\documentclass[onecolumn,11pt,tightenlines,superscriptaddress]{revtex4-1}

\usepackage{verbatim}
\usepackage{datetime}
\usepackage{graphicx}
\usepackage{bbm}
\usepackage{bm}
\usepackage{textcomp}
\usepackage{amssymb}
\usepackage{amsmath}

\begin{document}

\title{The evolving liaisons between the transaction networks\\of Bitcoin and its price dynamics}

\author{Alexandre Bovet}
\affiliation{ICTEAM, Université Catholique de Louvain, B-1348 Louvain-la-Neuve (Belgium)}
\affiliation{naXys, University of Namur, B-5000 Namur (Belgium)}
\author{Carlo Campajola}
\affiliation{Scuola Normale Superiore, I-56126 Pisa (Italy)}
\author{Francesco Mottes}
\affiliation{Physics Department, Università di Torino, I-10125 Torino and Istituto Nazionale di Fisica Nucleare (Italy)}
\author{Valerio Restocchi}
\affiliation{University of Southampton, SO171BJ Southampton (UK)}
\affiliation{The University of Edinburgh, EH89YL Edinburgh (UK)}
\author{Nicol\`o Vallarano}
\affiliation{IMT School for Advanced Studies Lucca, I-55100 Lucca (Italy)}
\author{Tiziano Squartini}
\email{tiziano.squartini@imtlucca.it}
\affiliation{IMT School for Advanced Studies Lucca, I-55100 Lucca (Italy)}
\author{Claudio J. Tessone}
\affiliation{URPP Social Networks and UZH Blockchain Center, University of Zurich, CH-8050 Z\"urich (Switzerland)}

\begin{abstract} 
Cryptocurrencies are distributed systems that allow exchanges of native tokens among participants, or the exchange of such tokens for fiat currencies in markets external to these public ledgers. The availability of their complete historical bookkeeping opens up the possibility of understanding the relationship between aggregated users' behaviour and the cryptocurrency pricing in exchange markets. This paper analyses the properties of the transaction network of Bitcoin. We consider four different representations of it, over a period of nine years since the Bitcoin creation and involving 16 million users and 283 million transactions. By analysing these networks, we show the existence of causal relationships between Bitcoin price movements and changes of its transaction network topology. Our results reveal the interplay between structural quantities, indicative of the collective behaviour of Bitcoin users, and price movements, showing that, during price drops, the system is characterised by a larger heterogeneity of nodes activity.
\end{abstract}

\maketitle

\section*{Introduction}

Cryptocurrencies are online payment networks in which the verification of transactions (and therefore the safeguard of system consistency) is decentralised and does not require a trusted third party, thanks to the clever combination of cryptographic technologies \cite{antonopoulos2017mastering}. Bitcoin is the first and most widely adopted cryptocurrency: it was created by the internet persona  Satoshi Nakamoto,  who  introduced it in 2008 with the release of a white paper \cite{nakamoto2008bitcoin}. More in detail, it consists of a decentralised peer-to-peer network to which users connect (by different means) in order to trade the account units of the system, i.e. \emph{bitcoins}. A transaction is included in a replicated database, the \emph{blockchain}, after having been validated by a network of miners, according to the consensus rules that are part of the Bitcoin protocol \cite{halaburda2016beyond,glaser2017pervasive}. A remarkable feature of the transaction-verification mechanism the Bitcoin system relies on, is that the entire transaction history (i.e. since the creation of the currency) is openly accessible, although in a pseudo-anonymised way. That is, users are not recorded directly but their pseudonyms, called \textit{addresses}, are.
Having access to the complete transaction history permits to investigate the inter-dependencies between the behaviours of Bitcoin users and the price dynamics which allows to shed light on the mechanisms governing complex out-of-equilibrium economic systems\cite{Arthur1999,Squartini:2013,Battiston2016}.

Since there is not a univocal representation of the Bitcoin transaction network, the method of choice is often adapted to the specific questions under study. For instance, Kondor \textit{et al.} \cite{Kondor:2014} consider the network of transactions between addresses, showing that the number of incoming transactions reflects the nodes wealth. Another example is provided by the work of Parino \textit{et al.} \cite{parino2018analysis} who investigate the network of  international Bitcoin flows, identifying  socio-economic factors that drive its adoption across countries. Also, several studies focus on the problem of de-anonymising Bitcoin users \cite{androulaki2013evaluating, harrigan2016unreasonable}, which boils down to cluster multiple Bitcoin addresses by common ownership \cite{meiklejohn2013fistful} following \textit{ad hoc} heuristics, in some cases achieving high precision. In order to provide yet another insightful Bitcoin network representation, we rely on these techniques in the following analysis, depicting the Bitcoin user network as accurately as possible.

This paper contributes to the literature by studying the structural properties of four different representations of the Bitcoin transaction network, i.e. the network of transactions between \emph{addresses} and between \emph{users}, at both the \emph{weekly} and \emph{daily} time scales. Moreover, we explore the causal relationships between the evolution of the aforementioned topological quantities and the Bitcoin price. 
To the best of our knowledge, ours is the first study of this nature. In fact, only the early analysis of \cite{kondor2014inferring} finds evidence of correlations between the components of the Bitcoin address network (identified via a Principal Component Analysis) and price. A different stream of research has focused on the price dynamics, ignoring the economic activity within the Bitcoin transaction network \cite{CHEAH201532,wheatley2019bitcoin}.

By thoroughly analysing the relations between the structural evolution of the transaction networks and the evolution of price, we show that the Bitcoin system underwent a clear change of regime after the bankruptcy of Mt.Gox on \date{2014-02-24}, i.e. the principal market exchange at the time, known to be performing market manipulation \cite{gandal2018price,mcmillan2014inside}.
Investigating the structural changes of the transaction networks allows us to gain insight into the feedback loops between network measures and price,
invisible when only the price dynamics is considered.
Among other results, we find that the moments of the out-degree distribution of the transaction network in its different representations shed light on the relationship between the heterogeneity of users behaviours and the Bitcoin price, potentially indicating the presence of herding behaviour.

\section*{Results}

\paragraph{Dataset description.} In the present work, four network representations of Bitcoin are analysed and compared: the network of transactions between addresses (\emph{address network} or AN, hereby) and inferred users (\emph{user network} or UN, hereby), at both a weekly (from Sundays to Sundays) and a daily frequency. The level of detail of our dataset is unprecedented. More specifically, our data consist of $16\,749\,939$ users and $304\,111\,529$ addresses, exchanging a total number of transactions amounting at $224\,620\,265$ and $283\,028\,575$, respectively. In terms of traded volume, the transactions between users and addresses amount at $3\,114\,359\,679$ and $4\,432\,597\,496$ bitcoins respectively.

From a topological point of view, both the AN and the UN are \emph{binary}, \emph{directed} networks, i.e. they are completely specified by their binary, asymmetric adjacency matrix $\mathbf{A}$. In the case of the AN network, the generic entry $a_{ij}^{(\Delta t)}$ is equal to 1 if at least one transaction between \emph{address} $i$ and \emph{address} $j$ takes place (i.e. bitcoins are transferred from \emph{address} $i$ to \emph{address} $j$) during the time window $\Delta t$ and 0 otherwise; in the case of the UN network, the entry $a_{ij}^{(\Delta t)}$ is equal to 1 if at least one transaction between \emph{user} $i$ and \emph{user} $j$ takes place (i.e. bitcoins are transferred from \emph{user} $i$ to \emph{user} $j$) during the time window $\Delta t$ and 0 otherwise. Generally speaking, a node in the UN representation (i.e. a ``user'') is a \emph{group of addresses}, opportunely gathered by implementing the heuristics described in the ``Methods'' section.

\begin{figure}[t!]
\includegraphics[width=\textwidth]{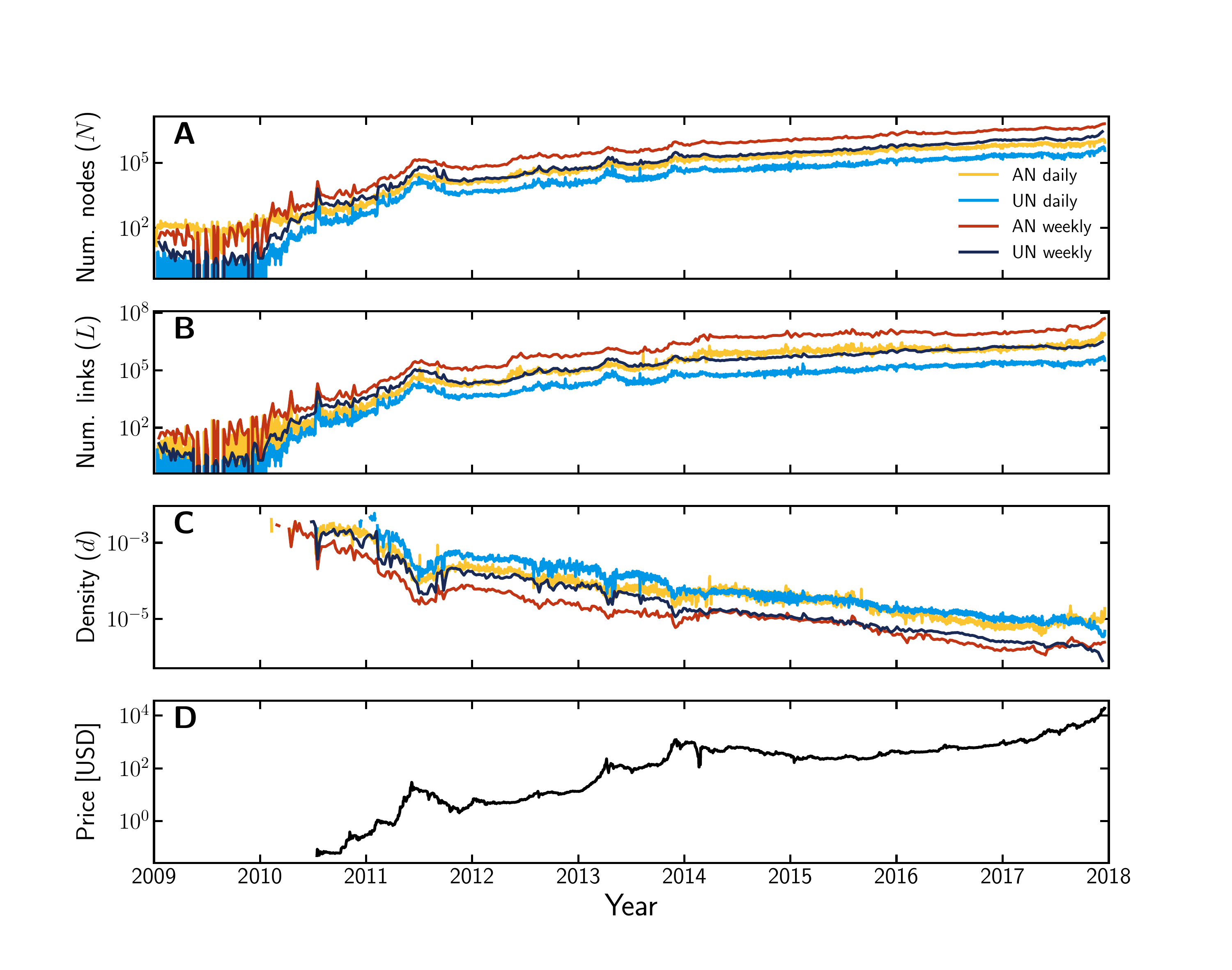}
\caption{\textbf{Evolution of basic statistics for the four Bitcoin network representations}. (\textbf{A}) number of nodes $N$, (\textbf{B}) the number of links $L$ and (\textbf{C}) the link density $d$ of the AN and the UN (reconstructed, see the "Methods" section for more detail), computed on a weekly and a daily basis, from \date{2009-1-9} to \date{2017-12-18}. The link density is computed only for networks with at least 500 nodes. The fourth panel (\textbf{D}) shows the evolution of the Bitcoin-US Dollar price, across the same period.}
\label{fig1}
\end{figure}

\paragraph{Evolution of the number of nodes, number of links and link density.} Figure \ref{fig1} shows the temporal evolution of the networks size and density, together with the evolution of the Bitcoin price. While both the total number of nodes $N$ (i.e. either users or addresses) and the total number of links $L$ between them are characterised by an overall rising trend, the link density
\begin{equation}\label{eq:1}
d=\frac{L}{N(N-1)}
\end{equation}
decreases with time: the Bitcoin network, in other words, becomes sparser regardless of the considered representation. It is worth noticing that the analysis of the system at different time scales does not reveal dramatically different behaviours for the inspected quantities.

\begin{figure}[t!]
\includegraphics[width=\textwidth]{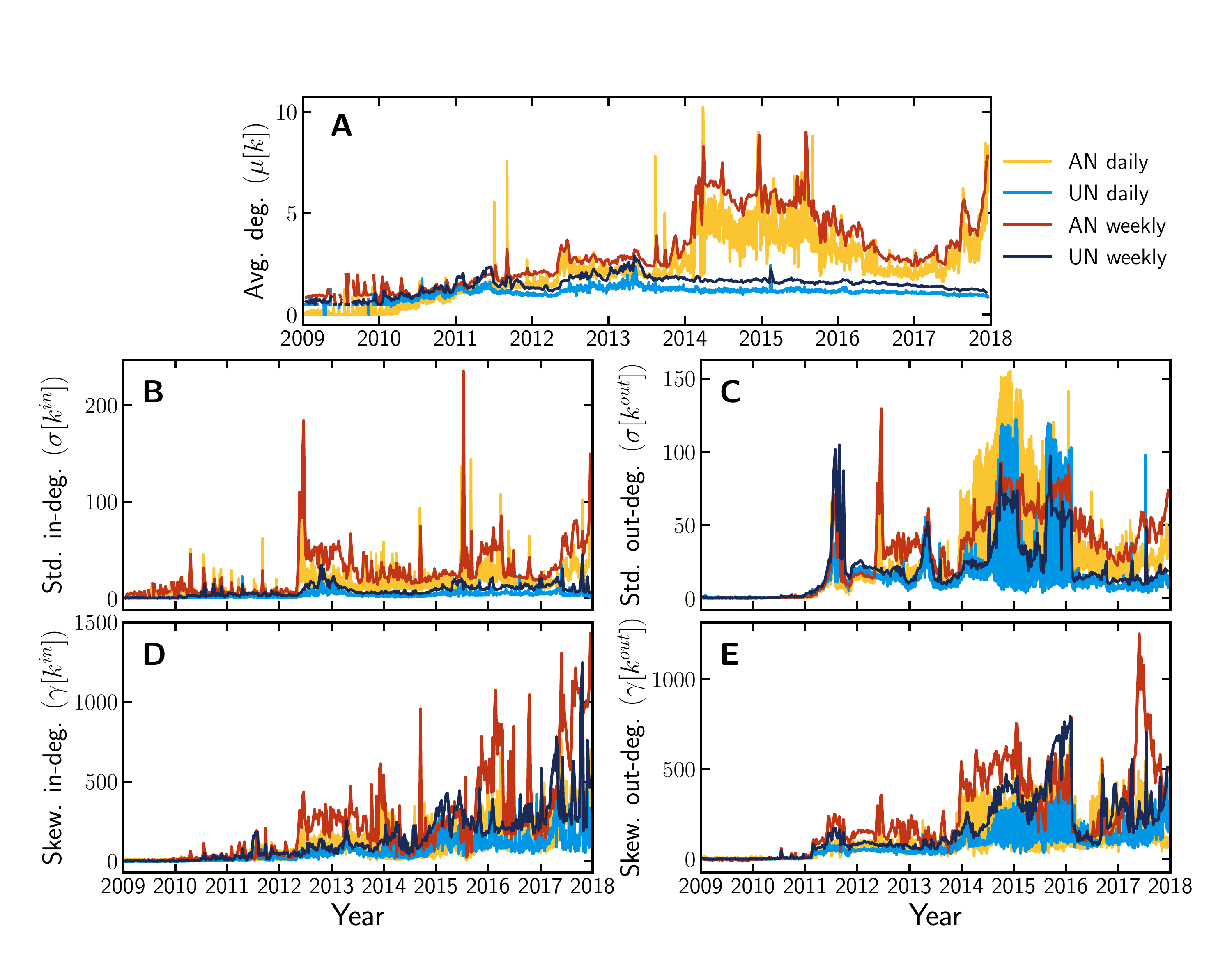}
\caption{\textbf{Evolution of moments of the in-degree and out-degree distributions}. In particular, evolution of (\textbf{A}) the average, $\bm{\mu}$, (\textbf{B},\textbf{C}) standard deviation, $\bm{\sigma}$ and (\textbf{D},\textbf{E}) skewness, $\bm{\gamma}$,  of the in-degree (\textbf{A},\textbf{B},\textbf{C}) and out-degree distributions (\textbf{A},\textbf{C},\textbf{E}). Differences between the AN and the UN representations of Bitcoin are visible: for example, while the average degree (\textbf{A}) of the UN is almost constant throughout the entire period we consider, its trend on the AN is characterised by peaks and oscillations. Different trends also characterise the evolution of the in-degrees and out-degrees standard deviation (\textbf{B},\textbf{C}): the latter is characterised by larger oscillations that reflect the larger heterogeneity of the payments, especially in the biennium 2014-2016. Moments of the in-degree distribution, on the other hand, tend to fluctuate less throughout the entire period, especially when the UN representation is considered. Similar considerations hold true when the skewness is analysed.}
\label{fig2}
\end{figure}

\paragraph{Degree distributions.} Given a directed network, the in-degree (respectively, out-degree) of a node $i$, noted $k^{in}_i$ (respectively, $k^{out}_i$), is the number of incoming (respectively, outgoing) links. The distributions of in-degrees and out-degrees are important measures to characterise the connectivity pattern of a network. The direction of links corresponds to that of bitcoins flow, consequently the in-degrees indicate the amount of bitcoins inflows (occurring when, for instance, bitcoins are purchased, a payment in bitcoins is received, a transfer between wallets is made, etc.), whereas out-degrees indicate the number of bitcoins outflows (occurring when, for instance, bitcoins are sold, a payment in bitcoins is made, a transfer between wallets is made, etc.). 

Upon noticing that the total number of links $L=\sum_i\sum_ja_{ij}=\sum_ik^{in}_i=\sum_ik^{out}_i$ in eq. \ref{eq:1} can be rewritten as
\begin{equation}\label{eq2}
d=\frac{\bm{\mu}\left[k\right]}{N-1}
\end{equation}
where $\bm{\mu}\left[k\right]={\sum_i k_i^{out}}/{N}={\sum_i k_i^{in}}/{N}$ is the average degree, i.e.  the average number of transactions per node (be it an address or a user), eq. \ref{eq2} suggests that, \emph{a priori}, two different contributions may lead to the observed, decreasing behaviour of $d$, namely an increment of the total number of nodes and a decrease of the average degree. Figure \ref{fig2}\textbf{A} shows the evolution of the average degree and allows the two contributions to be disentangled. Indeed, the evolution of the average degree of the user network is almost constant ($\bm{\mu}\left[k\right] \simeq 1.1$ for the daily UN), revealing that the decrease of the link density is clearly due to the increase of the network size: in more mathematical terms, $L=\mathcal{O}(N)$, inducing a decrease of the link density that evolves according to the functional form $d\sim N^{-1}$. Similarly, when the address network is considered, the dominating contribution comes from the evolution of $N$. However, Fig. \ref{fig2}\textbf{A} also shows a difference between the UN and the AN representations - the latter being characterised by peaks and oscillations, especially at the daily time scale - and a clear deviation between the two representations after 2014, suggesting that the Bitcoin network entered a different structural phase at this point.

Panels \textbf{B}-\textbf{E} from Fig. \ref{fig2} show the evolution of the heterogeneity of the in- and out-degree distributions measured by the standard deviation (Fig. \ref{fig2}\textbf{B}-\textbf{C}) and skewness (Fig. \ref{fig2}\textbf{D}-\textbf{E}) of the in- and out-degrees. It is possible to see that the out-degrees are characterised by a larger average volatility than in-degrees, whose values are instead more homogeneous throughout the entire period. In particular, when the AN representation is considered, the out-degrees standard deviation increases after 2014 and remains quite large until 2016. When the UN representation is considered, periods of large heterogeneity are observed from mid-2014 to 2015 and from mid-2015 to 2016. The skewness of the out-degree distributions (see Fig. \ref{fig2}\textbf{D}-\textbf{E}) increases as well in the same period.

\begin{figure}[t!]
\includegraphics[width=\textwidth]{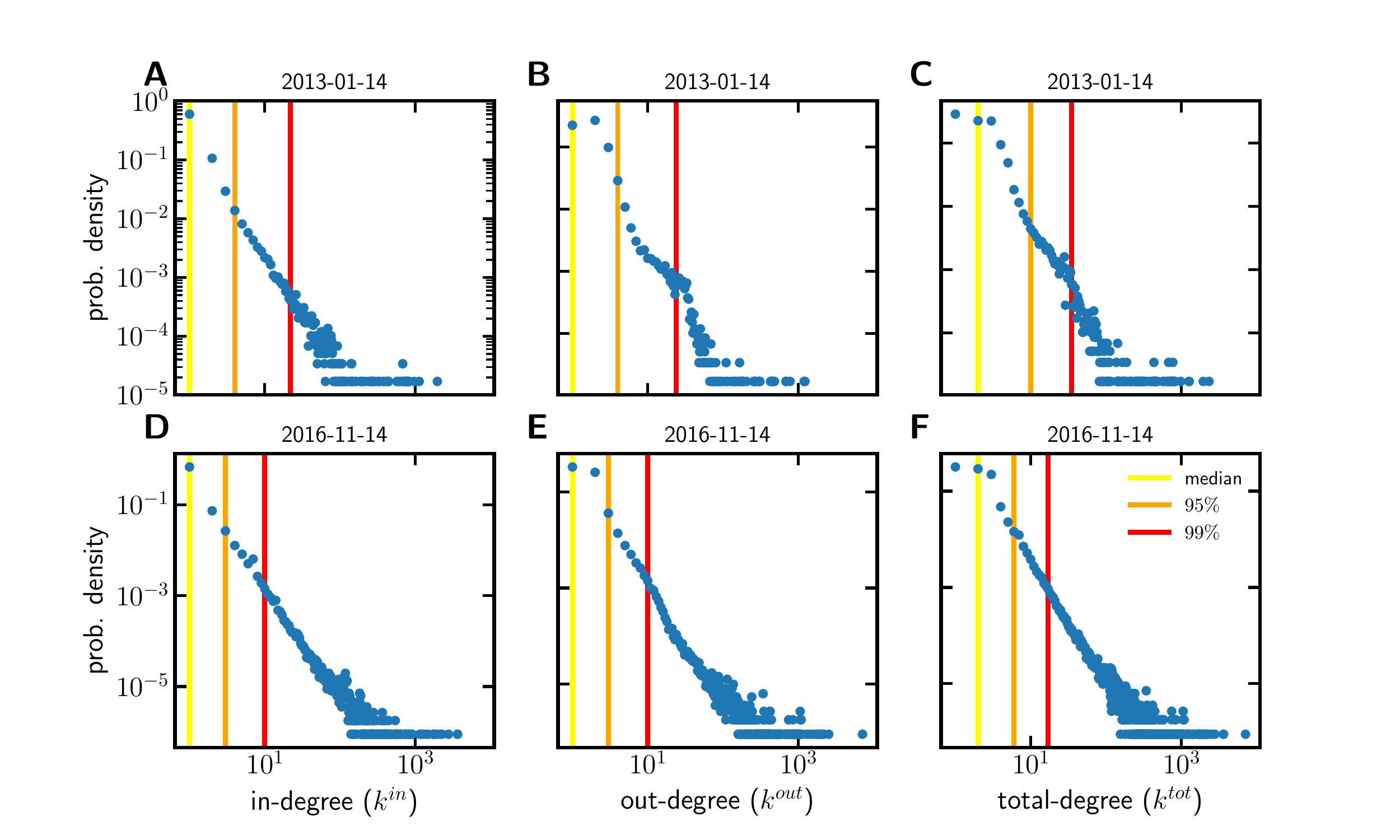}
\caption{\textbf{Probability density of the in-degree, out-degree and total degree distributions across our time interval}. Two snapshots of the in-, out- and total-degree distributions for the weekly UN, before and after 2014 (day \date{2013-01-14} and day \date{2016-11-14} for the top and bottom panels, respectively). These distributions are \emph{heavy-, right-tailed}, which suggests that most low-degree nodes coexist with few large hubs with thousands of incoming and outgoing connections. The median, 95th and 99th percentiles of the distributions are indicated in yellow, orange and red, respectively. By using the Kolmogorov-Smirnov test, we find that the hypothesis that the out-degrees are distributed according to a power-law cannot be rejected in $\simeq54\%$ of the snapshots before \date{2014-02-24} (i.e. the date of the Mt.Gox exchange market closure). This percentage drops to $\simeq26\%$ after that date. Conversely, the in-degree and total degree distribution can be considered compatible with a power-law distribution near half of the time.}
\label{fig3}
\end{figure}

\begin{figure}[t!]
\includegraphics[width=\textwidth]{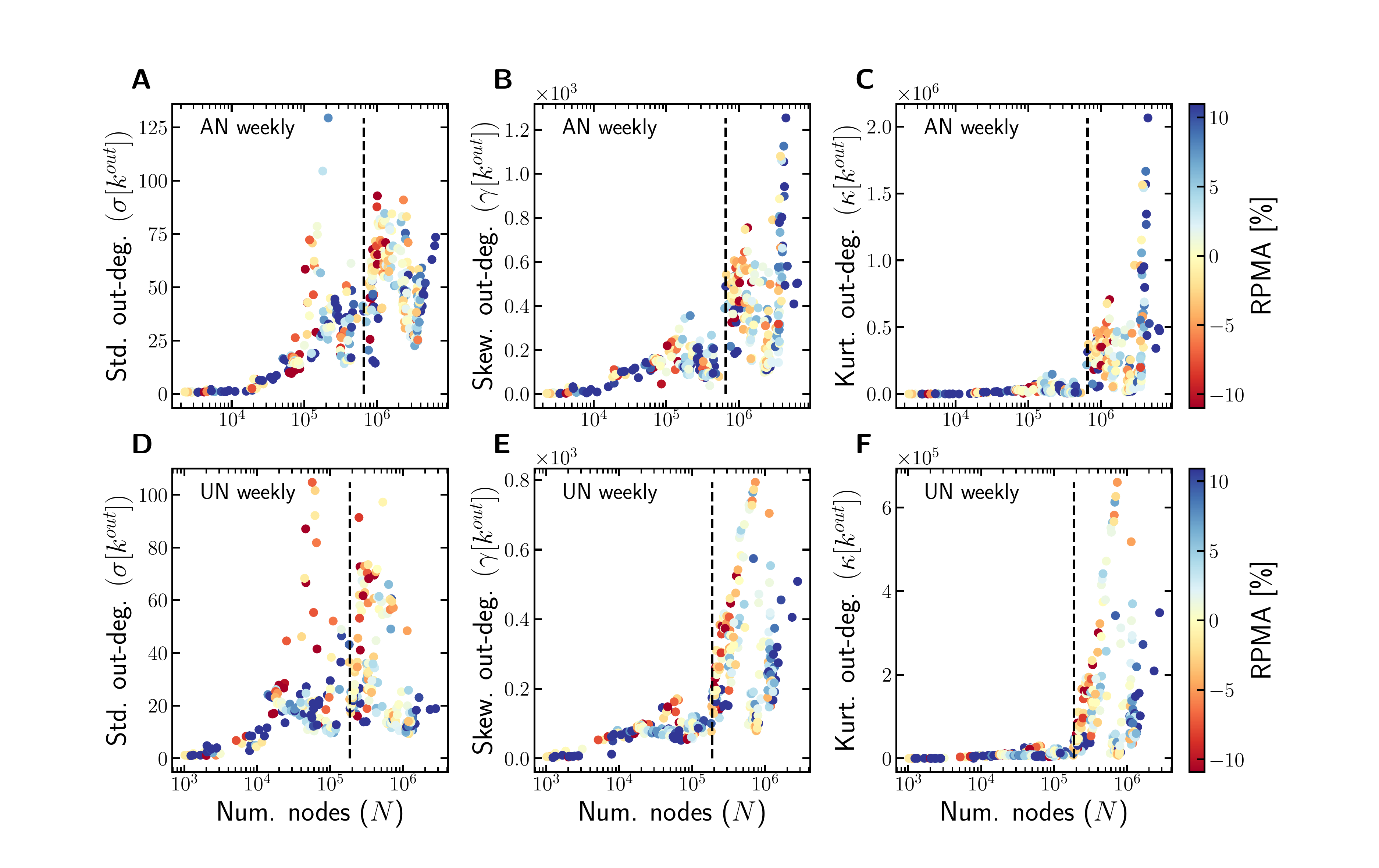}
\caption{\textbf{Correlation between the moments of the out-degree distribution, the number of nodes and the RPMA indicator}. This figure shows the relation between the moments of the network distributions and the number of nodes. Additionally, each dot representing an observation is coloured depending on the value of the Ratio between the current Price and its Moving Average (RPMA - see eq. (\ref{eq:log_returns})). These relations are shown for both the AN and the UN, in the top and bottom panels, respectively. The vertical, dashed line coincides with the bankruptcy of Mt.Gox in February 2014 and reveals two different regimes, ``before 2014'' and ``after 2014'': while the relation between the moments and the network size is well defined before 2014, it becomes less clear after 2014. The two main price drawdowns of 2011 and 2014 are also noticeable, being represented by the highest concentrations of points with negative RPMA values.}
\label{fig4}
\end{figure}

To show an example of the structural organisation of the Bitcoin transaction network, in Fig. \ref{fig3} we plot the distributions of in- and out-degrees for the weekly UN in two different periods. These distributions are \emph{heavy-, right-tailed}: this implies that a vast majority of nodes with low (in- and out-) degree coexists with few hubs with (tens of) thousands of connections; in some snapshots, a single hub with a degree of almost one order of magnitude larger than the second most connected node clearly emerges (see Fig. \ref{fig3}\textbf{E} and \ref{fig3}\textbf{F}).

To verify the functional form of the degree distributions depicted in Fig. \ref{fig3}, we employ the algorithm proposed in \cite{restocchi2019stylized} and based on the work by Bauke \cite{bauke2007parameter}. The algorithm we adopt is based on a double Kolmogorov-Smirnov statistical test whose results follow. First, the hypothesis that in-degrees and total degrees are distributed according to a power-law cannot be rejected for half of the considered snapshots: the percentage of times that the $p$-value is larger than 0.05 is $\simeq 45\%$ before \date{2014-02-24} and $\simeq 60\%$ after \date{2014-02-24} for in-degrees; for what concerns the total degrees, these percentages evolve from $\simeq 54\%$ to $\simeq 70\%$. However, these results change when out-degrees are considered. In fact, the percentage of times that the $p$-value is larger than 0.05 is $\simeq 54\%$ before \date{2014-02-24} and drops to $\simeq 26\%$ after \date{2014-02-24}. Although the out-degree distribution is heavy-tailed across the entire period, what emerges is that the failure of Mt.Gox had an impact on its \emph{functional form}.

An explanation of the observed evolution based on the functional form of the degree(s) distributions could call for a mechanism similar-in-spirit to the preferential attachment \cite{albert2002statistical} or undergoing a centrality-maximisation dynamics \cite{PhysRevE.84.056108}: each of the many users who are either new or extremely low-connected ``preferentially'' establishes a connection with a node that progressively grows to become one of the hubs. Such a node is likely to be an exchange market such as Mt.Gox, which takes part in transactions with (tens of) thousands of different users per week, while other (average) users typically trade once or twice per week (see Fig. \ref{fig2}\textbf{A}).

\begin{figure}[t!]
\centering
\includegraphics[width=\textwidth]{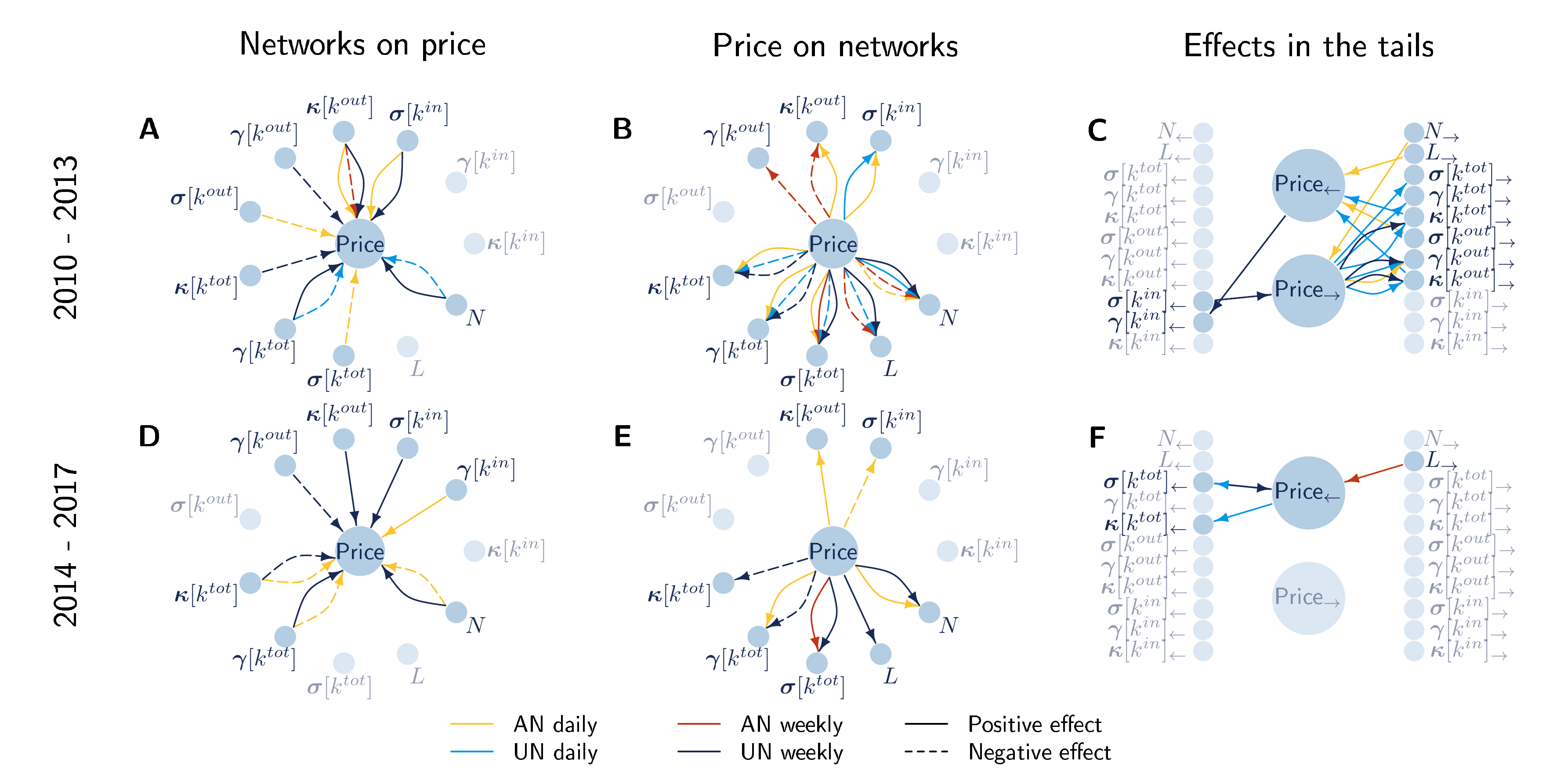}
\caption{\textbf{Analysis of the conditional Granger causality structure in the data}. The top and the bottom panels present the different causal relations for the periods 2010-2013 and 2014-2017, respectively. The left, centre and right columns show the effects of network properties on price, price on network properties and the restricted analysis to the tail values, respectively. 
(\textbf{A}) All higher moments of the out-degree distribution (i.e. the standard deviation, the skewness and the kurtosis) provide information on future price movements: the out-degrees standard deviation has a ``negative'' effect on the price, which means that an increasingly heterogeneous distribution of the number of payments causes a price drop and vice versa. (\textbf{B}) Upon looking at the opposite direction, price plays a major role in anticipating the moments of the distribution of total degrees, for all representations at all time scales. (\textbf{C}) The analysis of the Granger causality in tail in the period 2010-2013 identifies what can be anticipating a sudden crash in the market, i.e. what may cause extreme price movements in the left tail. (\textbf{D}, \textbf{E}) The causality structure of the weekly UN representation is largely unchanged; it drastically changes when considering the daily time scale: this result can be interpreted as a signature of increased market efficiency, leading to price unpredictability.}
\label{fig5}
\end{figure}

\paragraph{Relation between structural properties, network size and the Bitcoin price.} Here we analyse the relation between the structural evolution of both the AN and the UN transaction networks and the evolution of the Bitcoin price. Specifically, we employ the Ratio between the current Price and its Moving Average (RPMA), a commonly adopted technical indicator in financial analysis, to track the price dynamics. We consider both the weekly and daily logarithm of the RPMA and set a four-week and seven-day moving average, for the weekly and daily analysis respectively. Let $P_t$ be the closing price at week (or day) $t$, the RPMA at week (day) $t$  is computed as
\begin{equation}
\text{RPMA}_t= 100 \, \log_{10}\left(\frac{P_t}{\frac{1}{\tau}\sum_{s=t-1-\tau}^{t-1}P_{s}}\right)
\label{eq:log_returns}
\end{equation}
with $\tau=4$ for the weekly RPMA and $\tau=7$ for the daily RPMA. To understand the relation between the moments of the degree distribution and the network size, Fig. \ref{fig4} shows the normalised moments of order 2 (standard deviation), 3 (skewness) and 4 (kurtosis) of the weekly AN and UN networks as functions of the network size and of the RPMA (the same relations hold for the daily AN and UN network representations).

Figure \ref{fig4} reveals two regimes whose separation, indicated by a vertical dashed line, coincides with the bankruptcy of Mt.Gox in February 2014. Before 2014 the moments of the out-degree distributions tend to remain small and the relation between the moment value and the network size tends to be univocally defined; after 2014 their values tend to increase abruptly and the relation with network size is less clear suggesting that they may be diverging quantities. The divergence of moments after 2014 is also far more pronounced for the higher order moments (skewness, Fig. \ref{fig4}\textbf{B} and \textbf{E}, and kurtosis, Fig. \ref{fig4}\textbf{C} and \textbf{F}). The standard deviation (Fig. \ref{fig4}\textbf{A} and \textbf{D}), on the other hand, seems to be diverging mostly when the RPMA is negative. This relation with the RPMA is clearer on the UN network (Fig. \ref{fig4}\textbf{D}) than on the AN network (Fig. \ref{fig4}\textbf{A}), suggesting that aggregating the addresses may allow one to reveal clearer patterns.

Figure \ref{fig4}\textbf{D} shows that the out-degrees standard deviation of the weekly UN networks is characterised by an almost linear relation with the logarithm of the number of nodes, concerning the snapshots characterized by a positive RPMA, and from which two significant deviations appear approximately for $N\simeq 5\times10^4$ and $N\simeq 3\times10^5$, which coincide with price falls. These two deviations, which are also visible for the other moments, correspond to the two main drawdowns of the Bitcoin price (see Fig. \ref{fig1}\textbf{D}), the first of which starts after the bubble in mid-2011 and lasts until the beginning of 2012, whereas the second drawdown starts after the bankruptcy of Mt.Gox in early 2014 and continues approximately until 2016.

The period from 2016 to 2017 proves to be particularly interesting for a number of reasons. First, price reverts to a monotone, positive, trend. Second, the number of nodes steady increases (see Fig. \ref{fig1}\textbf{A} and \ref{fig1}\textbf{D}). Third, during these two years the standard deviation of the out-degree distributions decreases (see Fig. \ref{fig2}\textbf{C}). Fourth, as shown in Fig. \ref{fig4}\textbf{D}, the standard deviation of the out-degree distributions starts once again to show an almost linear relation with the logarithm of the number of nodes. Interestingly, the higher moments (skewness and kurtosis) do not revert to a more stable relation with the network size, suggesting that, although the RPMA has gone back to positive values and the standard deviation assumes stable values, the connectivity structure of the network has irreversibly changed after 2014, in particular for extremely connected nodes.

Since periods of negative returns are naturally identified with the drawdowns following the peak of the frequent bubbles occurring in the Bitcoin financial market, our analysis provides a structural characterisation of the latter. The dynamics taking place across the peak of a bubble can be loosely described as follows: when many users behave similarly, producing the same connectivity pattern (i.e. the degree distribution is sharper), a price surge is observed; as the peak is reached and crossed (i.e. the drawdown starts), individuals' connectivity tends to be more heterogeneous and the distribution widens. We interpret it as a network indicator of \emph{herding behaviour} on the Bitcoin market \cite{bouri2018herding}: during bubbles, Bitcoin users tend to have a very predictable connectivity and most of them only move bitcoins between the exchange and their wallet. Conversely, when a bubble bursts and the hoarding attitude declines, the everyday economic activities, represented on the network by heterogeneous degree distributions, are visible again.

\paragraph{Granger causality analysis in mean.} In order to gain further insight into the cross-market behaviour of Bitcoin, we carry out a causality analysis by using the Granger test \cite{granger1969investigating}, adopting a methodology frequently used in the literature \cite{gradojevic2014foreign, hong2009granger}. The simplest way to determine whether a first stochastic process $\{X_t\}$ causes a second stochastic process $\{Y_t\}$ is to carry out a \emph{bivariate causality test in mean}, aiming at verifying whether the response of a linear model including past information about $\{X_t\}$ significantly differs from the response of a model not including such information. A more refined version of the Granger test is \emph{multivariate}, which filters out effects of indirect causality (the bivariate test above may detect causality between a pair of variables because both are caused by a third, unaccounted variable) and also unveils causal relations that are hidden in the multivariate structure of the data (see  ``Methods'' section). We perform a multivariate Granger causality test on the two sub-samples considered above (i.e. 2010-2013 and 2014-2017). The network quantities that we consider are the number of nodes $N$, the number of links $L$ and the higher moments of the empirical (in- and out-) degree distributions (i.e. standard deviation $\bm{\sigma}$, skewness $\bm{\gamma}$ and kurtosis $\bm{\kappa}$ - for both the AN and UN representations, at both the daily and weekly time scales), which are then put in relation with the log-returns of the Bitcoin-US Dollar price, i.e.~$R_t=\log_{10}(P_t/P_{t-1})$.

The results concerning the causality structure are shown in Fig. \ref{fig5}. Specifically, in the period from 2010 to 2013 (panels \textbf{A}, \textbf{B}), out-degrees seem to play a major role in the price dynamics, i.e. all higher moments of the out-degree distribution anticipate future price movements. For instance, the kurtosis of the out-degree distribution has a predictive power at all time scales and for both representations (even if with opposite signs); this is no longer true when in-degrees are considered, as the only relevant contribution comes from their standard deviation (yet, providing information on future price movements at both the daily and the weekly time scales on the AN representation). Upon looking at the opposite direction (panel \textbf{B}), one realises that the price has several causal influences on the distribution of total degrees, for all representations at all time scales. Moreover, we detect an interesting relation between the number of nodes $N$ and  price: at the weekly timescale, in the UN representation, we observe a positive feedback loop between the two, whereas at the daily timescale a price increase would cause $N$ to increase, but an increase of $N$ would cause a price decrease. We observe these differences between fast and slow dynamics on the skewness of the total degrees too (see also later, i.e. the section devoted to the analysis of extreme events).

By comparing the market structure in the period 2010-2013 with the market structure in the period 2014-2017, shown in Fig. \ref{fig5}\textbf{D} and \ref{fig5}\textbf{E}, we find that the causality structure is remarkably consistent for the UN weekly data, but changes significantly at the daily time scale. More specifically, no causality relations are detectable when the daily UN representation in the period 2014-2017 is considered: this result can be interpreted as a signature of increased market efficiency at such a shorter time scale leading, in turn, to price unpredictability. Conversely, movements of structural quantities computed on the AN representation still provide information about price movements at the daily time scale.

\paragraph{Granger causality analysis in tail.} A third kind of causality analysis concerns the implementation of the so-called \emph{Granger causality in tail} as identified with the test by Hong et al. \cite{hong2009granger} and according to which an event is said to belong to the left (right) tail if it lies below the $10\%$ (above the $90\%$) empirical conditional quantile.

\begin{figure}
\centering
\includegraphics[width=\textwidth]{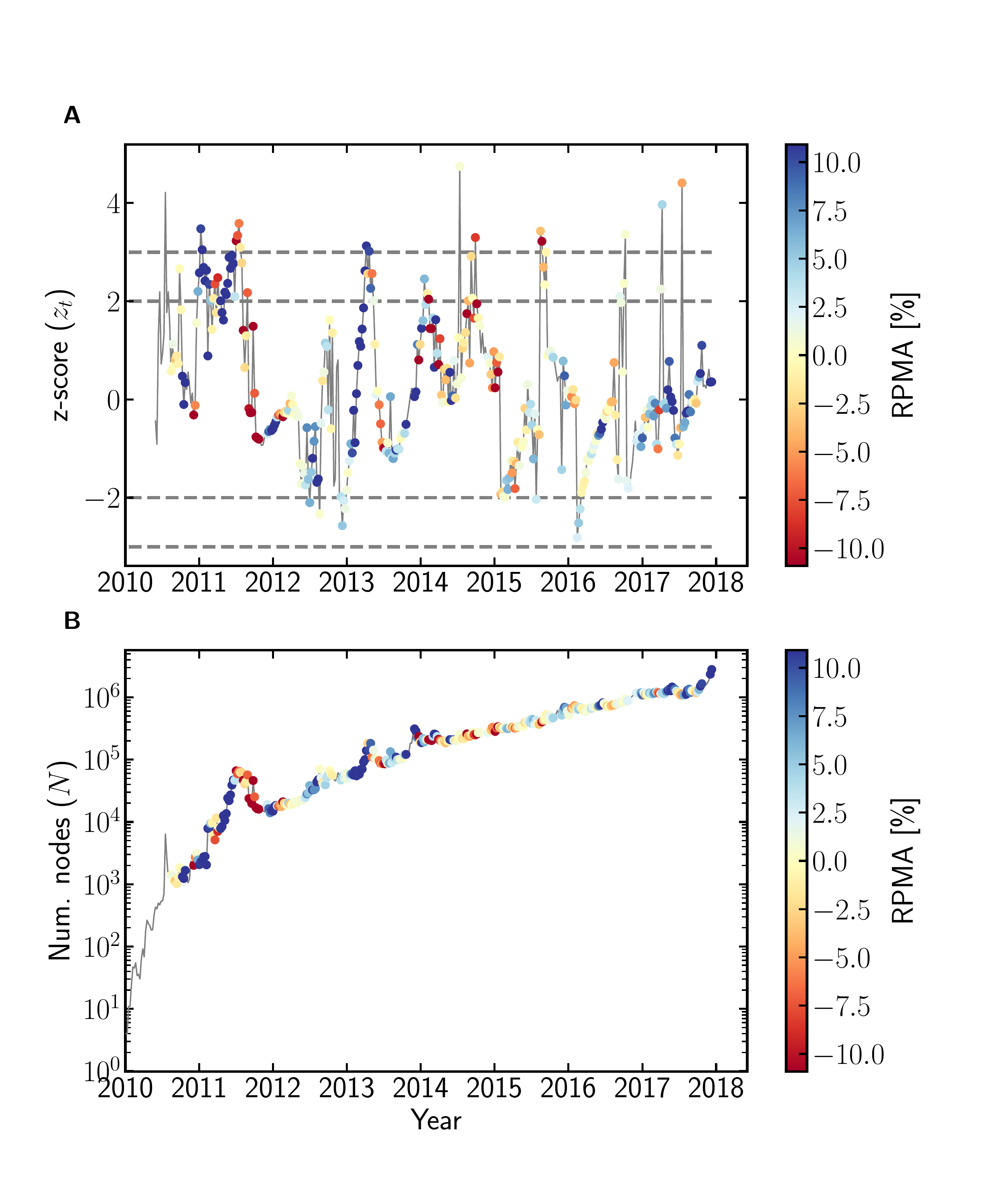}
\caption{\textbf{Evolution of the number of nodes and the $z$-score of the out-degrees standard deviation for the UN weekly representation}. Panel \textbf{A} shows the evolution of the temporal $z$-score of the out-degrees standard deviation. To compute the $z$-score, we consider mean and standard deviation over a period of one year before each snapshot. During the period from 2010 to 2013 it is possible to see that the $z$-score of the out-degrees standard deviation grows together with price (dark blue dots represent significantly large price changes). Drawdowns, instead, appear as periods during which the $\bm{\sigma}[k^{out}]$ $z$-score decreases. Moreover, our results reveals peaks between 2015 and 2016, suggesting that topological indicators provide evidence of ongoing structural changes, which are otherwise missed when only purely financial indicators as RPMA are employed. Interestingly, during year 2017 price surges correspond again to an increase of the $z$-score of the out-degrees standard deviation. Panel \textbf{B} shows the evolution of the number of nodes.}
\label{fig6}
\end{figure}

The results obtained from this analysis,  shown in Fig. \ref{fig5}\textbf{C} and \ref{fig5}\textbf{F}, suggest that the tail causality structure differs significantly from the one in mean, mainly because of the different time scale at which causality is detected. Specifically, many tail causal relations are found only at the daily scale, rather than at the weekly scale: in particular, as $N$ increases, the Bitcoin price increases as well. Conversely, as $\bm{\sigma}[k^{out}]$ increases, the Bitcoin price decreases. This kind of analysis is particularly relevant to identify what can be anticipating a sudden crash in the market. It is, in fact, enough to inspect what may cause extreme price movements in the left tail. For example, as the out-degree distribution becomes more heterogeneous (at the daily scale), a sudden price drop is likely to follow. Before 2014 this effect is indeed present, both on the AN and the UN representations, for both the second and the fourth moment of the out- and total degree distributions.

Combining the causality in mean with the causality in tail allows a better understanding of how the Bitcoin transaction networks evolve with respect to the exchange price; particularly relevant are two feedback loops on multiple timescales: (1) the loop involving the number of nodes in the UN representation; (2) the loop involving the out-degrees kurtosis in the UN representation. The first, as already mentioned, has differently signed effects whether one looks at the daily or weekly timescale and tells the story of a market that grows in size, leading to higher price valuations and further growth of the number of users, in a slow feedback mechanism that takes place over several weeks. Conversely, at a faster timescale of few days lag, a rise in the number of active nodes leads the price down, possibly explained by a ``return to normal'' after unjustified short-term price movements caused by the increase in popularity. The second loop occurs when one joins the information from the Granger causality with the analysis of Fig. \ref{fig4}. Specifically, an increased $\bm{\kappa}\left[k^{out}\right]$ in the UN representation leads, on average, to a price increase on the weekly timescale, which as we have seen causes $N$ to increase. Looking at Fig. \ref{fig4} we see that $N$ and $\bm{\kappa}\left[k^{out}\right]$ are positively correlated, thus closing the feedback loop on the UN weekly scale into a self-reinforcing mechanism (albeit weaker than the first as we do not see a direct causation from the price to the $\bm{\kappa}\left[k^{out}\right]$). When looking at the results from the tail causality tests, though, we see that an abnormal increase in $\bm{\kappa}\left[k^{out}\right]$ causes the price to crash on the UN daily representation. This shows that a possible explanation to the frequent crashes occurring before 2014 is given by this second feedback loop reaching an unsustainable growth rate, leading to violent mean reversions in the financial market.

\paragraph{Temporal analysis.}
To better understand the dynamics of the relation between network structure and price, we consider one of the most informative moments of the out-degree distribution, the observed value of the standard deviation of the out-degrees $\bm{\sigma}_t[k^{out}]$ in the snapshot at time $t$. To achieve this, we compute the $z$-score by normalising this value as follows
\begin{equation}
z_t=\frac{\bm{\sigma}_t[k^{out}]-\bm{m}_t\big[\bm{\sigma}[k^{out}] \big]}{\bm{s}_t\big[{\bm{\sigma}[k^{out}]}\big]}
\end{equation}
where $\bm{m}_t\big[\cdot \big]$ and $\bm{s}_t\big[\cdot \big]$ represent the average and the standard deviation of the argument over the preceding $\tau$ time steps (see the ``Methods'' section for the definition of $\tau$). The following results are obtained by using an averaging period of one year (similar results were obtained with both shorter and longer averaging periods).

When considering the UN representation, the temporal evolution of $z_t$, shown in Fig. \ref{fig6} together with the price RPMA, suggests that the dynamics of $z_t$ highlights, and in some cases anticipates, the critical periods previously discussed. In particular, when considering the 2010-2013 time window, price surges correspond to positive increments of the out-degrees standard deviation; similarly, drawdowns are characterised by $z_t$ drops. It is also possible to notice that our analysis reveals peaks in 2015 and 2016, suggesting that topological indicators may provide evidence of ongoing structural changes that are missed when just the price is monitored; indeed, according to the standard price-based analysis, the 2014-2016 time window corresponds to a unique drawdown. The year 2017 constitutes a period of particular interest as the behaviour of the system changes quite drastically with respect to the period 2014-2016: specifically, price surges correspond again to an increase of the $z$-score of the out-degrees standard deviation.

Finally, we would like to stress that the results above are not in contradiction with the results shown in Fig. \ref{fig4}. Indeed, whereas Fig. \ref{fig4} depicts the absolute values of the out-degrees standard deviation across the entire temporal window we consider in this paper (i.e. 2010-2017), the points in Fig. \ref{fig6} result from temporal comparisons with the previous year values. In other words, an increase of $z_t$ does not necessarily reflect an increase of the $\bm{\sigma}[k^{out}]$ value \emph{on a global scale}, but only means that the out-degrees standard deviation rises \emph{with respect to the average computed over the values of the previous year}.

\section*{Discussion}

Several years after the date of its release, Bitcoin has shown to be still able to attract an increasing number of users, both because of speculative reasons\cite{gandal2018price,chu2015statistical,hayes2017cryptocurrency} and the trust of early adopters in the potential of this innovative technology \cite{bohme2015bitcoin,gervais2014bitcoin}. In fact, the number of users and transactions have steadily increased, while the Bitcoin market value and its volatility have witnessed remarkable bursts, leading to gold-rush-like bubbles and unexpected price crashes \cite{Garcia:2014,wheatley2018bitcoin,gerlach2018dissection}.

The structure and dynamics of the Bitcoin transaction networks are still poorly understood. For this reason, in this paper we consider four Bitcoin network representations and analyse the interplay of macroscopic structural properties with the Bitcoin price movements over the first 10 years of its evolution (i.e. by considering the  period ranging from 2009 to 2017). More specifically, we focus on aggregate network properties and analyse the evolution of their distribution moments.

Our results show that it is possible to recognise two different phases of the Bitcoin system, one between 2010 and 2013, and the second one from 2014 to 2017. The event that had a dramatic impact on the system can be identified in the bankruptcy of Mt.Gox, the incumbent market exchange at the time, also known to be performing market manipulation\cite{gandal2018price}. Bitcoin prices were hit tremendously by the incident and only in January 2017 the Bitcoin-US Dollar exchange rate once again surpassed the $\$1.000$ threshold, thanks to a renovated interest in cryptocurrencies. Throughout 2017, the Bitcoin price surged from $\$963$ to $\$14.112$ at an average daily rate of $0.7\%$ price increase. 

Before 2014, the moments of the out-degree distribution obey an almost linear relation with the logarithm of the number of nodes, with deviations corresponding to important price drops. After 2014, although the relation between the standard deviation of the out-degree distribution and the network size is still well defined, the higher moments of the out-degree distribution become more unstable, indicating a change in the structural dynamics of the network.
With respect to this, out-degrees have been found to be particularly informative properties, as their heterogeneity rises during periods of price decline and vice versa. Such a result is further consolidated by a Granger causality analysis in tail, revealing that - during the 2010-2013 period - an increase of the out-degrees standard deviation \emph{causes} a price decline.

Our results also suggest an explanation for the price dynamics displayed during the early stages of Bitcoin: during \emph{bull markets} (i.e. periods in which the price continuously increases), an ever increasing number of traders is attracted to the market, thus causing the price to rise even more. As a consequence, hubs, which usually represent exchanges, get a large number of connections over the course of the weeks, coming from a large number of users performing only few transactions. This self-reinforcing mechanism gets to a point where, on shorter timescales, it becomes convenient for speculators to suddenly revert the trend, causing a market crash on the daily scale. This effect is somehow mitigated in the later stages of the market (i.e. 2014-2017), when no causal relations are found on the daily timescale and the price trajectory is much smoother.

Overall, our results indicate that the existing interplay between structural quantities and price movements indeed suggests an alternative way to gain insight into the system evolution. One of such insights is the observation that the dynamics of the degree heterogeneity, driven by the collective behaviour of those who are part of the system, provides a powerful network indicator of \emph{herding behaviour} on the Bitcoin market \cite{bouri2018herding}. 

To conclude, our research shows that these closed economies serve as an extraordinary information source to analyse economic and financial processes ranging from the micro to the macro scales: we have demonstrated that there exists a co-evolving liaison between the economic activity within the Bitcoin system, measured by very simple topological indicators in the network of transactions, and the exchange price dynamics. Future research may examine more intricate topological traits and the role of specific actors in the price formation mechanisms.

\section*{Materials and Methods}

The Bitcoin system relies on a blockchain, i.e. a decentralised public ledger which records all transactions of bitcoins among users. A transaction is a set of \emph{input} and \emph{output addresses}, involved in the exchange of a specific amount of bitcoins.
The unspent \emph{output addresses} (i.e. not yet recorded on the ledger as \emph{input addresses}) can be claimed, and therefore spent, only by the owner of the corresponding cryptographic key; this is the reason why we speak of \emph{pseudonimity}: an observer of the blockchain can see all unspent addresses but he cannot link them to actual owners.

By using the publicly available raw data on transactions, we analysed two network representations of Bitcoin: the network of transactions between addresses (AN) and the network of transactions between users (UN), on both a weekly and a daily basis, in order to exploit different levels of granularity. In what follows we will describe the algorithm we have implemented to define such representations.

\begin{description}
\item[Address Network (AN)]
The address network is the simplest network that can be constructed from the blockchain records, and is a directed, weighted graph in which each node represents an address. Direction and weight of links are provided by the input-output relationships defining the transactions recorded on the blockchain. The only free parameter is represented by the temporal window chosen for the data aggregation. In this paper, we chose to aggregate these data both daily and weekly. While the former is the shortest scale one can consider that still guarantees that the resulting network is connected, the weekly aggregation follows from the natural cadence of financial markets and the calibration of the Bitcoin mining difficulty which gets updated once every two weeks \cite{nakamoto2008bitcoin}.

\item[Users Network (UN)]
Since the same owner may control several addresses, based on the transactions network one can derive a \emph{users network} in which the nodes are clusters of addresses (i.e. the nodes of the transactions network) and links are recovered upon aggregation of the links in the AN. Such clusters are aggregated by implementing different \textit{heuristics}, i.e. sets of rules which take advantage of the implementation of the Bitcoin protocol in order to assess whether different addresses belong to the same user. Below we describe the heuristics we employed, which we have derived from the state-of-the-art literature \cite{androulaki2013evaluating,tasca2018evolution,harrigan2016unreasonable,ron2013quantitative}.

\begin{description}
\item[Multi-input heuristics] This heuristic is generally believed to be the safest one for clustering addresses, and is based on the assumption that, if two (or more) addresses are part of the input of the same transaction, then they are controlled by the same user. The key idea behind this heuristics is that, in order to produce a transaction, the private keys of all addresses must be accessible to the creator.

\item[Change-address identification heuristics] Transaction outputs must be fully spent upon re-utlisation. Hence, the transaction creator usually controls also one of the output addresses. Specifically, we consider that if an output address is new and the amount transferred to it is lower than all the inputs, then it must belong to the input user.
\end{description}
\end{description}

Whenever the \textit{users network} is mentioned in the paper, we refer to the representation obtained by clustering the AN nodes according to a combination the two heuristics above. As evinced by the different figures, smoother trends are observed for the UN representation as a consequence of using these heuristics to aggregate addresses into ``users''. 

We also stress that users can use different wallets that are not necessarily linked together by transactions. As a consequence, the user network we obtained should not be considered as a perfect representation of the real network of users, but rather a best-effort approximation which enables us to group addresses while minimising the presence of false positives. Once the addresses are grouped into ``users'', we build the network in the following way: any two nodes $i$ and $j$ are connected via a directed edge from $i$ to $j$ if at least one transaction from one of the addresses defining $i$ to one of the addresses defining $j$ occurs at the considered time scale.

\subsection*{Granger causality in mean}

The simplest causality test we performed is a \emph{bivariate causality test in mean}, as proposed in the original paper by Granger\cite{granger1969investigating}: it aims at verifying whether a linear model that includes past information about $X$ has a residual sum of squares (RSS) that is significantly different from the RSS of a model that does not include such information. In more detail, given two stochastic processes $X_t$ and $Y_t$ the two models are defined as
\begin{equation}
Y_t=\sum_{k=1}^\tau \alpha_k Y_{t-k}+\epsilon_t^Y,
\end{equation}
and
\begin{equation}
Y_t=\sum_{k=1}^\tau \beta_k Y_{t-k}+\sum_{k=1}^\tau \beta_k X_{t-k}+\epsilon_t^{XY}
\end{equation}
where $\tau$ is the maximum lag considered and $\{\epsilon_t\}$ is a series of i.i.d. standardised Gaussian random variables. A standard $F$-test is then performed to test whether the variance of $\epsilon^{XY}$ is significantly different from the variance of $\epsilon^X$: if this is the case, the null hypothesis of no causality can be rejected. As a limitation, this bivariate test may detect causality between a pair of variables simply because both are caused by a third, unaccounted one.

\subsection*{Multivariate Granger causality}

A more refined, \emph{multivariate} test aims at filtering out effects of indirect causality as well as bringing to light causal relations that are hidden in the multivariate structure of the data. In this case, all the $N^{ts}$ time series, represented in tensor form as $\mathbf{A}_t$, are fitted into a VAR model as
\begin{equation}
\mathbf{A}_t=\sum_{k=1}^\tau \mathbf{B}_k\textbf{A}_{t-1}+\Xi_t+\mathbf{c}
\end{equation}
where $\{\Xi_t\}$ is a series of multivariate standardised Gaussian random variables and $\mathbf{c}$ is a vector of constants. The formulation above allows one to test whether variable $i$ ``Granger-causes'' variable $j$ conditionally on the presence of all the other variables, by fitting the reduced model
\begin{equation}
\mathbf{A}_t^{(-i)}=\sum_{k=1}^\tau \mathbf{B}_k^{(-i)}\mathbf{A}_{t-1}^{(-i)}+\Xi_t^{(-i)}+\mathbf{d},
\end{equation}
where $\mathbf{A}^{(-i)}$ is the data matrix after removing column $i$. Performing the $F$-test between the RSS of $\Xi^{(-i)}_j$ and $\Xi_j$. The choice of the maximum lag $\tau$ is arbitrary, and in this work we choose to consider $\tau=4$ for weekly data and $\tau=7$ for daily data, corresponding to a temporal window of a month and a week, respectively. Regarding the daily data, we choose $\tau=7$ because we do not want effects to be overlapping with those of the weekly scale. As for the weekly data, we argue that events further apart than one month would not be meaningfully related in realistic terms, and consequently choose $\tau = 4$ for this timescale.

\subsection*{Granger causality in tail}

The third kind of Granger causality we consider is the one between extreme events, measured on distributions \emph{tails}. Here, we implement the test proposed by Hong et al.\cite{hong2009granger} for events outside the $10\%-90\%$ conditional quantiles range. Briefly speaking, an event is assumed to belong to the left tail of a given distribution if its probability is below the $10\%$ conditional quantile; analogously, an event is assumed to belong to the right tail if its probability exceeds the $90\%$ conditional quantile. Quantiles are computed by employing the method by Davis\cite{davis2016verification} on a trailing rolling window of one month for the daily data and of two months for the weekly data.

Given two time series $X_t$ and $Y_t$, two binary series marking the occurrence of tail events $Z_t$ and $W_t$ are constructed as follows. Call $\sigma(X_{t-1})$ the past history of $X_t$ up to time $t$ and $q^X$ the quantile, conditional on $\sigma(X_{t-1})$, beyond which an event is considered to be in the $\alpha$-tail, that is $\mathbb{P}(X_t>q_t^X\vert\bm{\sigma}(X_{t-1}))=\alpha$, where $\mathbb{P}$ is the empirical conditional CDF for $X_t$. Then
\begin{equation}
Z_t=\mathbb{I}(X_t>q_t^X)\:\:\:\text{and}\:\:\:W_t=\mathbb{I}(Y_t>q_t^Y)
\end{equation}
where $\mathbb{I}$ is an indicator function. The test devised by Hong et al.\cite{hong2009granger} is based on the fact that the spectral density of a linear model between $Z$ and $W$ is diagonal under the null hypothesis of no causality. A test statistic can, then, be defined to check how likely the null hypothesis is:
\begin{equation}
Q(M)=\frac{\left[T\sum_{l=1}^{T-1} K^2(l/M)\hat{\rho}^2(l)-C_{T}(M)\right]}{\sqrt{D_T(M)}}
\end{equation}
where $T$ is the length of the two considered time series, and $K(z)$ is a kernel function with bandwidth $M$ (in our case, a Daniell kernel has been employed, i.e. $K(z)={\sin(\pi\:z)}/{\pi\:z}$). The bandwidth parameter's role is key to set the scale at which causality between tail events is inspected: a broader kernel would be more sensitive to causality relations, while a narrower bandwidth would have increased power at the expense of sensitivity. $\hat{\rho}(l)$ is the empirical cross-correlation function between $Z$ and $W$ at lag $l$; $C_T(M)$ and $D_T(M)$ are a centring and a scaling constant respectively, depending on the choice of the kernel and making $Q(M)$ distributed as a standard normal with zero mean and unit variance under the null hypothesis of no causality.

As mentioned earlier, conditional quantiles are estimated via the non-parametric methodology proposed by Davis\cite{davis2016verification}. Briefly speaking, consider the estimate $\hat{q}^X_t$ of the $\alpha$-quantile that is based on the past $M$ observations: that would be the $\alpha M$-th observation sorted in magnitude. 

If the series presents heteroscedasticity and apparent non-stationarity the estimate will be biased and a simple data-driven correction overcomes this issue: instead of taking $\hat{q}^X_t$ as the conditional quantile estimate, the expression $q^X_t=\hat{q}^X_t+\phi(\check{y}_{t-1}-(1-\alpha))$ has to be considered, where $\check{y}_t=\frac{1}{t}\sum_{t'=1}^{t}\mathbb{I}(X_{t'}>q^X_{t'})$ and $\phi$ is a free parameter tuning the feedback mechanism.

To account for multiple hypothesis testing and reduce the number of false positives, a False Discovery Rate (FDR) \cite{benjamini1995controlling} at $5\%$ is used in every set of Granger Causality tests.

\subsection*{$z$-score}

The $z$-score of a generic quantity $X$ is defined as $Z_X={X-\bm{\mu}_X}/\bm{\sigma}_X$, and represents a standardised variable that measures the difference between the observed and $\bm{\mu}_X$ (the expected value of $X$) in units of standard deviation $\bm{\sigma}_X$. If $X$ is normally distributed under the considered model, values within $z=\pm 1, 2, 3$ would occur with a probability $p(z) \cong 68\%, 95\%, 99\%$ respectively. If the observed value of $X$ induces a largely positive (negative) value of $Z_X$ then the quantity $X$ is said to be over (under)-represented in the data and, as such, not explainable only by the topological constraints defining the null model.

Alternatively, the $z$-score can be also computed by considering the sample averages $\bm{m}[X]$ and standard deviation $\bm{s}[X]$. In this case, it reads $z_X=\frac{X-\bm{m}[X]}{\bm{s}[X]}$.

\bibliographystyle{ScienceAdvances}
\bibliography{Bibliography}

\noindent\textbf{Acknowledgements:} All authors are grateful to Alberto Antonioni, Eugenio Valdano and the IMT School for Advanced Studies Lucca for the organisation of the workshop ``Complexity72h'' during which this project was started.
\textbf{Funding:} A.B. acknowledges financial supports by the ARC (Action de Recherche Concert\'{e}e) on Mining and Optimisation of Big Data Models funded by the Wallonia-Brussels Federation and the Swiss National Science Foundation (Fellowship P300P2\_177793).
C.C. acknowledges Scuola Normale Superiore for financial support of the project SNS18\_A\_TANTARI. V.R. was sponsored by the U.S. Army Research Laboratory and the U.K. Ministry of Defence under Agreement Number W911NF-16-3-0001. The views and conclusions contained in this document are those of the authors and should not be interpreted as representing the official policies, either expressed or implied, of the U.S. Army Research Laboratory, the U.S. Government, the U.K. Ministry of Defence or the U.K. Government. The U.S. and U.K. Governments are authorised to reproduce and distribute reprints for Government purposes notwithstanding any copyright notation herein. C.J.T. acknowledges financial support of the University of Zurich through the University Research Priority Program on Social Networks.
\textbf{Author Contributions:} A.B., C.C., F.M., V.R., N.V. performed the analysis. A.B., C.C., F.M., V.R., N.V., T.S. and C.J.T designed the research. All authors wrote, reviewed and approved the manuscript.
\textbf{Competing Interests:} The authors declare that they have no competing financial interests.
\textbf{Data and materials availability:} Data concerning the entire Bitcoin transaction history is publicly available at the address https://www.blockchain.com/ (last access: 4 July 2019). Moreover, by syncronizing the desktop client Bitcoin Core (available at https://bitcoin.org/en/download) anybody can also have a local copy of the entire transaction history.
\end{document}